\documentclass[final,5p,times,twocolumn]{elsarticle}
\biboptions{numbers,sort&compress}

\usepackage{amssymb}
\usepackage{lipsum}

\usepackage{amsmath}
\usepackage{amssymb}
\usepackage{graphicx}
\usepackage{dcolumn}
\usepackage{bm}
\usepackage{braket}
\usepackage{mathrsfs}
\usepackage{booktabs}
\usepackage{multirow}
\usepackage{url}
\usepackage{physics}
\usepackage{subfigure}
\usepackage[colorlinks=true,linkcolor=blue, filecolor=blue,urlcolor=blue,citecolor=blue]{hyperref}
\usepackage{color}
\usepackage{cuted}

\journal{Physics Letters B}

\begin{document}
	
	\begin{frontmatter}

\title
{Quantum computing for extracting nuclear resonances}

\author[first]{Hantao Zhang}
\author[second]{Dong Bai}
\author[first,third]{Zhongzhou Ren\corref{cor1}}
\ead{zren@tongji.edu.cn}
\cortext[cor1]{Corresponding author}

\affiliation[first]{organization={School of Physics Science and Engineering},
	addressline={Tongji University}, 
	city={Shanghai},
	postcode={200092}, 
	country={China}}
\affiliation[second]{organization={College of Mechanics and Engineering Science},
	addressline={Hohai University}, 
	city={Nanjing},
	postcode={211100}, 
	country={China}}
\affiliation[third]{organization={Key Laboratory of Advanced Micro-Structure Materials},
	addressline={Tongji University}, 
	city={Shanghai},
	postcode={200092}, 
	country={China}}

%
%
%

\begin{abstract}
Quantum computing has been increasingly applied in nuclear physics. In this work, we combine quantum computing with the complex scaling method to address the resonance problem. Due to the non-Hermiticity introduced by complex scaling, standard quantum computing cannot solve for complex eigenvalues directly. Therefore, it is necessary to embed the non-Hermitian operator into a larger dimensional unitary operator. Additionally, for the case of two basis vectors, we improve the traditional direct measurement method and optimize the quantum circuit. Ultimately, using the $\alpha+\alpha$ system as an example, we obtain the complex eigenenergies from the quantum computer that are consistent with those obtained from direct Hamiltonian diagonalization.

\end{abstract}


\end{frontmatter}

%
%
%
%

%
%
%
%
%
%
%
%
\section{introduction}
Quantum computing represents a revolutionary paradigm in computational science, leveraging the principles of quantum mechanics to process information in fundamentally new ways. As an emerging field at the intersection of quantum mechanics and computer science, quantum computing has been increasingly applied in various scientific domains, such as computational chemistry, nuclear physics, and beyond. Central to this advancement are the concepts of quantum entanglement and quantum information, which enable unprecedented computational power and efficiency. 
 The entanglement properties of hadrons and  nuclei are also increasingly gaining attention in nuclear physics \cite{Ho_2016,PhysRevD.95.114008,PhysRevD.98.054007,PhysRevLett.122.102001,Tu_2020,BEANE2021168581,ISKANDER2020135948,Kruppa_2021,doi:10.1142/S0217751X21502055,PhysRevD.103.065007,PhysRevD.104.L031503,PhysRevC.106.024303,PhysRevA.107.012415,PhysRevC.103.034325,PhysRevD.104.074014,PhysRevD.106.L031501,Klco_2022,Bai:2022hfv,Johnson_2023,EHLERS2023169290,Tichai_2023,Pazy_2023,Bulgac_2023,PhysRevA.105.062449,PhysRevC.107.044318,PhysRevC.105.014307,Bai:2023rkc,Robin_2023,Gu_2023,Sun_2023,Bai:2023tey,Bai:2023hrz,Bai:2024omg}. Traditional classical computers struggle with the exponential scaling of computational resources needed to accurately model and simulate nuclear systems, particularly when dealing with many-body problems. However, quantum computing, with its natural capacity to represent and process quantum information, offers a promising solution to these challenges. Quantum algorithms, such as the variational quantum eigensolver (VQE) \cite{osti_1623945,Peruzzo_2014,McClean_2016,PhysRevLett.120.210501,RevModPhys.94.015004,PhysRevC.104.024305} and its various variants \cite{Higgott_2019,McArdle_2019,Yuan_2019,Grimsley_2019,PhysRevResearch.2.043140,Stokes_2020,Gomes_2021,PRXQuantum.2.020310,PhysRevC.105.064317,Koczor_2022} and quantum phase estimation (QPE) \cite{kitaev1995quantummeasurementsabelianstabilizer,Nielsen_Chuang_2010}, leverage the unique characteristics of quantum computing to provide new pathways for solving eigenvalue problems, fundamental in determining the energy spectra of nuclear systems. The potential for quantum computers to perform these calculations more efficiently than classical computers opens up new possibilities for advancing our understanding of nuclear physics, as well as other complex quantum systems.

One specific area where quantum computing can make a significant impact is in the study of resonance. Resonant states are unstable and have complex energy values, whose imaginary parts indicate the half-life. The study of resonance is critical for understanding nuclear reactions and decay processes. In this work, we combine quantum computing with the complex scaling  method (CSM) \cite{Aguilar1971ACO,Balslev1971SpectralPO,10.1143/PTP.116.1,Myo:2014ypa,PhysRevC.89.034322,Zhang:2022rfa,Zhang:2023dzn,Myo:2023btg}, a powerful technique for dealing with resonant states.  The complex scaling  method is based on a  complex transformation which converts the wave function of resonant state to square-integrable and thus the resonances can be treated with bound-state techniques. However, the non-Hermiticity introduced by such complex transformation presents challenges for conventional quantum algorithms, which typically require Hermitian operators. To overcome this, one can embed the non-Hermitian operator into a larger dimensional unitary operator \cite{SK2013,10.1063/5.0040477}, enabling the use of quantum algorithms to find complex eigenvalues. Additionally, we optimize quantum circuits and improve the direct measurement method to enhance the efficiency and accuracy. By applying our approach to the $\alpha+\alpha$ system, a well-studied example in nuclear physics, we demonstrate that the results obtained from the quantum computer are consistent with those from direct diagonalization methods.

{\color{black}In addition to the complex potential introduced by complex scaling , other types of complex potentials have also been extensively investigated and already successfully applied in nuclear physics. For instance, the conventional optical potential is utilized in nuclear scattering problems to describe the absorptive effects of nuclei on nucleons, and the  effective short-range complex potential is employed in studying  antinucleon-nucleon scattering to elucidate the intricate dynamics of nucleon-antinucleon pair annihilation.
Such complex potentials also pose challenges in solving for complex eigenvalues with quantum computing. For instance, if we use the trap method  \cite{Zhang:2024vmz,Zhang:2024mot,Zhang:2024ykg} to obtain the scattering phase shift in a system with complex potentials, we need to find the eigenvalues corresponding to the complex Hamiltonian \cite{Zhang:2024ykg}.}

The rest of the article is organized as follows: In Sec.\ref{Theoretical Formalism},  we present the complex scaling method and discuss how to construct the quantum circuits to obtain the eigenenergies of resonances by applying the quantum computer.
In Sec.\ref{Numerical Results} the numerical results of resonance of $\alpha+\alpha$ system are
presented and discussed. Sec.\ref{Conclusions} summarizes the article. 
Some details are presented in the Appendix.

\section{theoretical formalism}
\label{Theoretical Formalism}
Through complex scaling method the complex scaled Hamiltonian  is obtained from  a similarity transformation,
\begin{equation}
	\begin{aligned}
		H(\textbf{r},\theta)=U(\theta)H(\textbf{r})U(\theta)^{-1},
	\end{aligned}
\end{equation}
where  $U(\theta)$ is defined as
\begin{equation}
	\begin{aligned}
		U(\theta)f(\textbf{r})=\exp(i\dfrac{3}{2}\theta)f(\textbf{r}\exp(i\theta)),  0<\theta<\pi/2,
	\end{aligned}
\end{equation}
with this new Hamiltonian the corresponding Schr{\"o}dinger equation is written as,

\begin{equation}
	\begin{aligned}
		H(\textbf{r},\theta)\Psi(\textbf{r},\theta)=E(\theta)\Psi(\textbf{r},\theta).
	\end{aligned}
\end{equation}

Based on the ABC theorem \cite{Aguilar1971ACO,Balslev1971SpectralPO}, after the complex transformation one can obtain the following conclusions, (1) the eigenvalue of a bound state of $H$ is still an eigenvalue of $ H(\theta)$; (2) a resonance pole  $E-i\Gamma/2$ is an eigenvalue of $H(\theta)$; (3)
the continuum spectrum of $H$ is rotated into the fourth quadrant of complex energy plane by the angle $2\theta$. With the help of CSM the wave functions of resonant states become square integrable, therefore one can expand the wave function over a set basis functions.
After selecting  a proper orthogonal basis set $\{\phi_i(r)\}$, the complex scaled Hamiltonian can be expressed as a second quantization form,
\begin{equation}
	\begin{aligned}\label{}
		&H_{\theta}=\sum_{i,j}C_{i,j}(\theta)a_i^{\dagger}a_j^{},
	\end{aligned}
\end{equation}
where  $a^{\dagger}$ and $a$ are fermionic creation and annihilation operators. The elements of matrix $C$ can be obtained by,
\begin{equation}
	\begin{aligned}\label{}
		&C_{i,j}(\theta)=\int \phi_i^{}(r)(\eta^{-2}T+V(r\eta))\phi_j(r)d^3r,
	\end{aligned}
\end{equation}
where $\eta=se^{i\theta}$, here $s$ is chosen as 1.

The fermionic creation and annihilation operators can be expressed through the Jordan-Wigner transformation \cite{Jordan1928berDP},
\begin{equation}
	\begin{aligned}\label{}
		&a_j^{\dagger}=\dfrac{1}{2}(X_j-iY_j)\otimes Z_{j-1}^{\mathscr{D}},\\
				&a_j^{}=\dfrac{1}{2}(X_j+iY_j)\otimes Z_{j-1}^{\mathscr{D}},
	\end{aligned}
\end{equation}
where $X$, $Y$, and $Z$ are Pauli operators, $Z_{j-1}^{\mathscr{D}}$  is defined as
\begin{equation}
	\begin{aligned}\label{}
		&Z_{j-1}^{\mathscr{D}}=Z_{j-1}\otimes Z_{j-2}\otimes \cdots \otimes Z_0.
	\end{aligned}
\end{equation}

According to Jordan-Wigner transformation, the complex scaled Hamiltonian will be further transformed into Pauli operators as,

\begin{equation}
	\begin{aligned}\label{PauliSum}
		&H_{\theta}=\sum_{i}\beta_iP_i,
	\end{aligned}
\end{equation}
where $\beta_i$ represents the complex coefficients, and $P_i$ represents a $k$-local tensor product of Pauli operators,
where $k \leq n$ and $n$ is the size of the basis set.

\begin{figure}[htbp] 
	\centering
	{\includegraphics[width=0.5\textwidth]{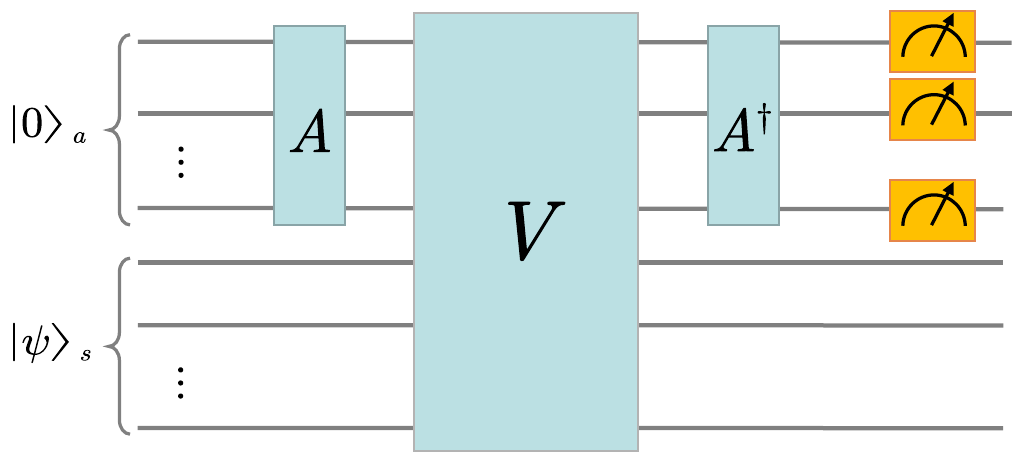} }
	\caption{The quantum circuit for direct measurement method. A and V gates are constructed based on
		the coefficients and operators in Eq.(\ref{generalcircuit}). The system qubits’ state and ancilla qubits’ state are initialized as $\ket{0}$  and $\ket{\psi}$ respectively. The input eigenvector can be constructed by introducing a variational wave function ansatz
		based on unitary coupled-cluster (UCC) theory \cite{McClean_2016,PhysRevA.95.020501}.}	\label{bigcircuit} 
\end{figure}

The above process is the same as the conventional Hamiltonian derivation in quantum computing for bound states. Here for resonance calculations, to make the Hamiltonian more
compatible with the direct measurement method, we rewrite Eq.(\ref{PauliSum}) as,
\begin{equation}
	\begin{aligned}\label{}
		&\alpha_i=|\beta_i|,\ V_i=\dfrac{\beta_i}{|\beta_i|}P_i.
	\end{aligned}
\end{equation}

The effects of the A and V gates in the circuit can be formulated as,

\begin{equation}
	\begin{aligned}\label{generalcircuit}
		&A\ket{0}_a=\sum_{i=0}^{2^{n_a}-1}\sqrt{\dfrac{\alpha_i}{\alpha}}\ket{i}_a,\  \sum_{i=0}^{2^{n_a}-1}\alpha_i=\alpha,\\&
		V\ket{i}_a\ket{\psi}_s=\ket{i}_aV\ket{\psi}_s.
		\end{aligned}
\end{equation}

If we denote $W=(A^{\dagger}\otimes I^{\otimes n_s})V(A\otimes I^{\otimes n_s})$, then the result of the whole circuit acting on the initial state can be expressed as,
\begin{equation}
	\begin{aligned}\label{}
		&W \ket{0}_a\ket{\psi}_s=\dfrac{Ee^{i\phi}}{\alpha}\ket{0}_a\ket{\psi}_s+\ket{\Psi}_{\bot},
	\end{aligned}
\end{equation}
where $Ee^{i\phi}$ ($E>0$) is the eigenvalue and  the ancilla qubits's state of $\ket{\Psi}_{\bot}$ is perpendicular to $\ket{0}_a$.

\section{numerical results}
\label{Numerical Results}

\begin{figure}[htbp] 
	\centering
	{\includegraphics[width=0.5\textwidth]{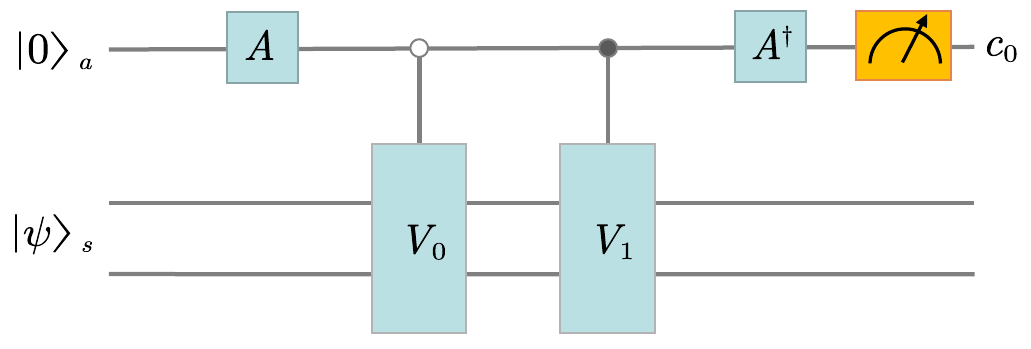} }
	\caption{The quantum circuit for two basis wave functions and one ancilla qubit. A and V gates are constructed based on
		the coefficients and operators in Eq.(\ref{generalcircuit}). The system qubits’ state and ancilla qubits’ state are initialized as $\ket{0}$  and $\ket{\psi}$ respectively.}	\label{circuit} 
\end{figure}

Due to noise and errors in quantum computing, quantum circuits generally need to be optimized to reduce the number of qubits and basic quantum gates. In this work, we construct a circuit which contains one ancilla qubit and two system qubits (i.e. two basis functions) as shown in Fig.\ref{circuit}. For this specific situation, it can be proven that the measurement results are independent on the input eigenvector. Therefore we can further simplify and transform the original circuit into one that contains only two qubits and fewer quantum gates. {\color{black}By measuring two circuits with the same structure but different quantum gate parameters, we can obtain two possibilities $p_1$ and $p_2$, allowing us to simultaneously determine the real and imaginary parts of the resonant energy $Ee^{i\phi}$.}  More details can be found in the Appendix.

Below, we will use quantum computing to study the $\alpha-\alpha$ resonance. In this work we employ both the two-body toy model \cite{BUCK1977246} and the microscopic, Tohsaki-Horiuchi-Schuck-Röpke (THSR) wave function method \cite{Zhou_2013, Zhang:2022rfa,Zhang:2023dzn} to describe $\alpha+\alpha$ system. Firstly, we study the $\alpha+\alpha$ toy model with quantum computing. The two-body interaction is modeled by a Gaussian potential, and the Coulomb potential is represented using the error function form,
\begin{equation}
	\begin{aligned}\label{}
		&V_N(r)=V_0 \exp(-\kappa r^2),\\&
		V_C(r)=\dfrac{Z_1Z_2e^2}{r}\erf(\beta r).
	\end{aligned}
\end{equation}
where the parameters $V_0=-122.6225$ MeV, $\kappa=0.22$fm${}^{-2}$, and $\beta=0.75$fm${}^{-1}$ are taken from \cite{BUCK1977246}. Here we use the harmonic oscillator basis $\ket{nL}$ to expand the wave function,
where $\hbar\omega=\dfrac{\hbar^2}{m_Nb_0^2}$ and $b=\dfrac{b_0}{\sqrt{2}}$ with $b_0$ being 1.36 fm.



Since the effect of Pauli blocking is not considered in this interaction, both S-wave and D-wave will have redundant bound states. Of course, this issue does not exist for the microscopic THSR wave function method. In addition, we can exclude these redundant states by projecting the model space using the orthogonality condition model (OCM) \cite{10.1143/PTP.40.893}, or we can ensure that the trial wave function is orthogonal to the redundant states by introducing the generalized coherent states, a method proposed by Myo et
al.  and already successfully applied in the study of $\alpha+\alpha$ system \cite{Myo:2024vrs}. Here, for simplicity, we only choose the G-wave resonance that has no redundant states. The resonant energy of G-wave is calculated to be $11.7823-1.7867i$ MeV.

\begin{figure}[htbp] 
	\centering
	{\includegraphics[width=0.5\textwidth]{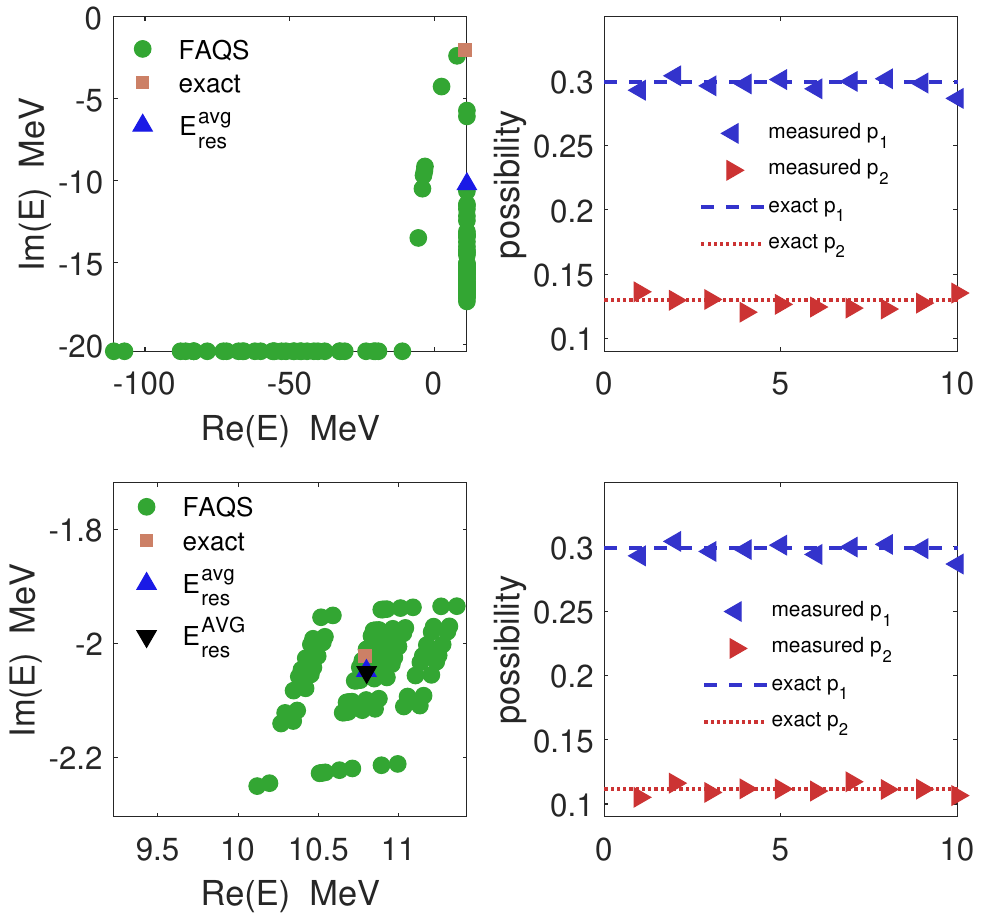} }
	\caption{Top left panel: The resonant energies obtained from quantum simulation FAQS with $\theta=25.4343^{\circ}$ and global scaling $g=1$ applied to the Hamiltonian. Green circles represent the results from FAQS, orange squares indicate the exact values obtained through diagonalization, and blue triangle represents the result obtained by averaging measured possibilities.Top right panel: Blue left triangles and red right triangles represent the results of FAQS for $p_1$ and $p_2$ respectively. The dashed and dotted lines indicate the exact values of $p_1$ and $p_2$, respectively. Bottom left panel: Similar with top left panel but with $g=25-10i$. The black down-triangle represents the result obtained by directly averaging 100 data points. Bottom right panel: Similar with top left panel but with $g=25-10i$. }	\label{FAS_oringinal_GS} 
\end{figure}

Generally, when the CSM is employed to extract the resonance, the optimal complex scaling angle $\theta$ can be determined by the pause point, where the rate of change of the complex energy with respect to the angle $\theta$ is minimized. Practically, when using different types of basis functions or different numbers of the same type of basis function, the corresponding optimal angles are different from each other. Moreover, with an insufficient number of basis functions, it may be impossible to determine a global optimal angle. Therefore, in this example with only two basis functions (i.e., two system qubits) are involved in quantum computation, we do not strictly adhere to the choice of $\theta$. Instead, we focus on demonstrating the effectiveness of our method itself.
{\color{black} In the following part we only select the angle $\theta=25.4343^{\circ}$ corresponding to the eigenvalue $10.79-2.022i$ MeV closest to the true eigenenergy as an example to validate the effectiveness of our approach.}

 \begin{figure}[htbp] 
	\centering
	{\includegraphics[width=0.5\textwidth]{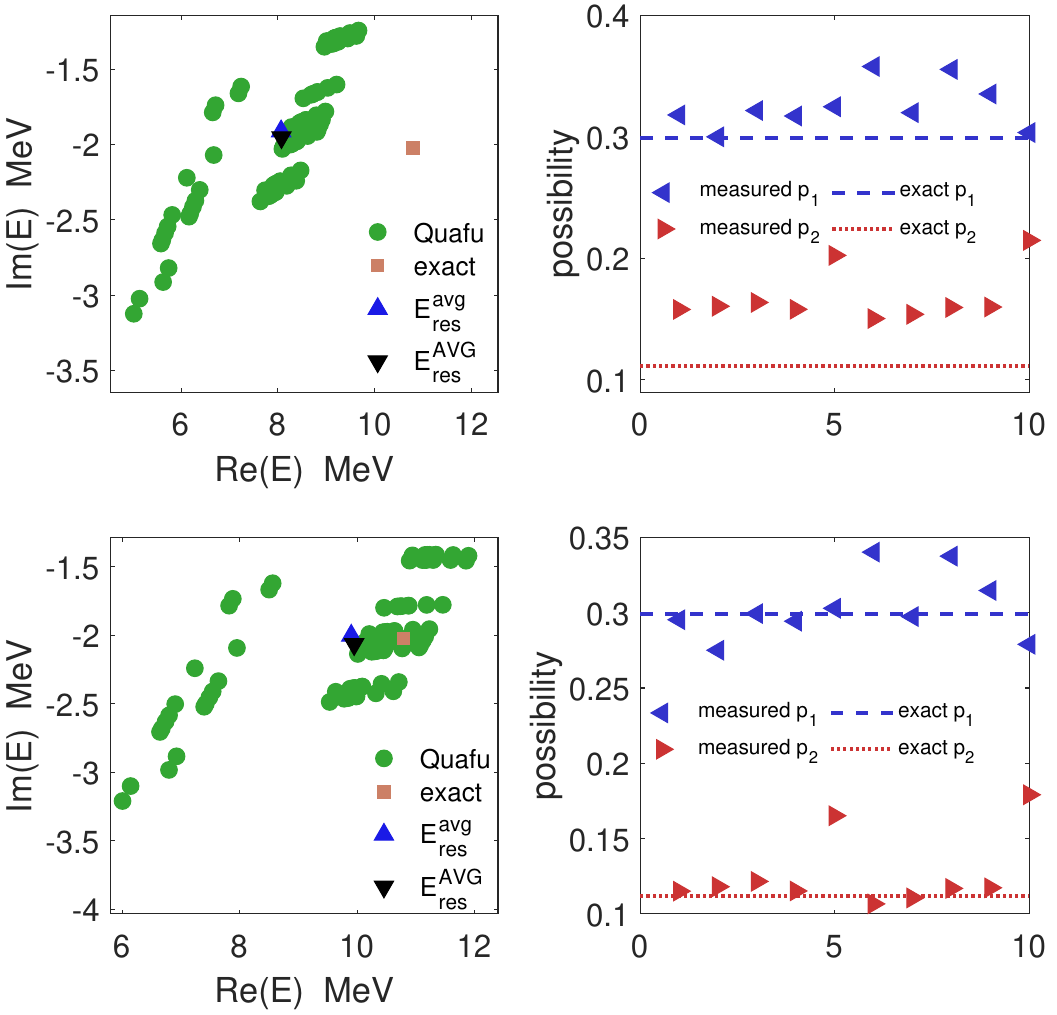} }
	\caption{The resonance state energies obtained from quantum computer Quafu-Dongling with  $\theta=25.4343^{\circ}$ and global scaling $g=25-10i$. The top two figures correspond to Dongling's original measurements, while the bottom two figures show the results after readout error mitigation. The green dots represent the  results obtained from Quafu-Dongling, the orange squares represent the exact values, and the blue upward triangles and black downward triangles correspond to the results from averaging possibilities and averaging all eigenvalues, respectively. The blue left triangles and red right triangles denote the measurement possibilities  $p_1$ and $p_2$, with the blue dashed and red dotted lines representing the corresponding exact values.}	
	\label{quafuDongling} 
\end{figure}

{\color{black}Before performing calculations on a real quantum computer, we first use a quantum simulator to measure qubits and obtain complex eigenvalues. The simulator used here is the Origin quantum's full-amplitude quantum simulator (FAQS) and the number of measurements is 8192 ($2^{13}$). 
The results obtained from FAQS are displayed in Fig.\ref{FAS_oringinal_GS}. The top left and right panels correspond to the original Hamiltonian. Green circles represent the results from FAQS, orange squares indicate the exact values obtained through direct Hamiltonian diagonalization, and blue triangle represents the result obtained by averaging measured possibilities. The blue left triangles and red right triangles represent the measurement possibilities $p_1$ and $p_2$, with the blue dashed and red dotted lines denoting the corresponding exact values.
It can be seen that the possibility results obtained from FAQS are relatively accurate with small errors. However, the dispersion in the extracted eigenvalues is too large to reliably determine the final eigenenergy. Moreover, as seen in top left panel in Fig.\ref{FAS_oringinal_GS}, even averaging the possibilities does not yield an acceptable result. This is due to the high sensitivity of the eigenvalues to the measurement possibilities $p_1$ and $p_2$, resulting in very weak robustness of the solution. Even a small error in quantum computing can lead to significant distinct in the final eigenvalues.
Such bad news may completely prevent us from recovering the eigenvalues from the measurement possibilities.}
\begin{figure}[htbp] 
	\centering
	{\includegraphics[width=0.4\textwidth]{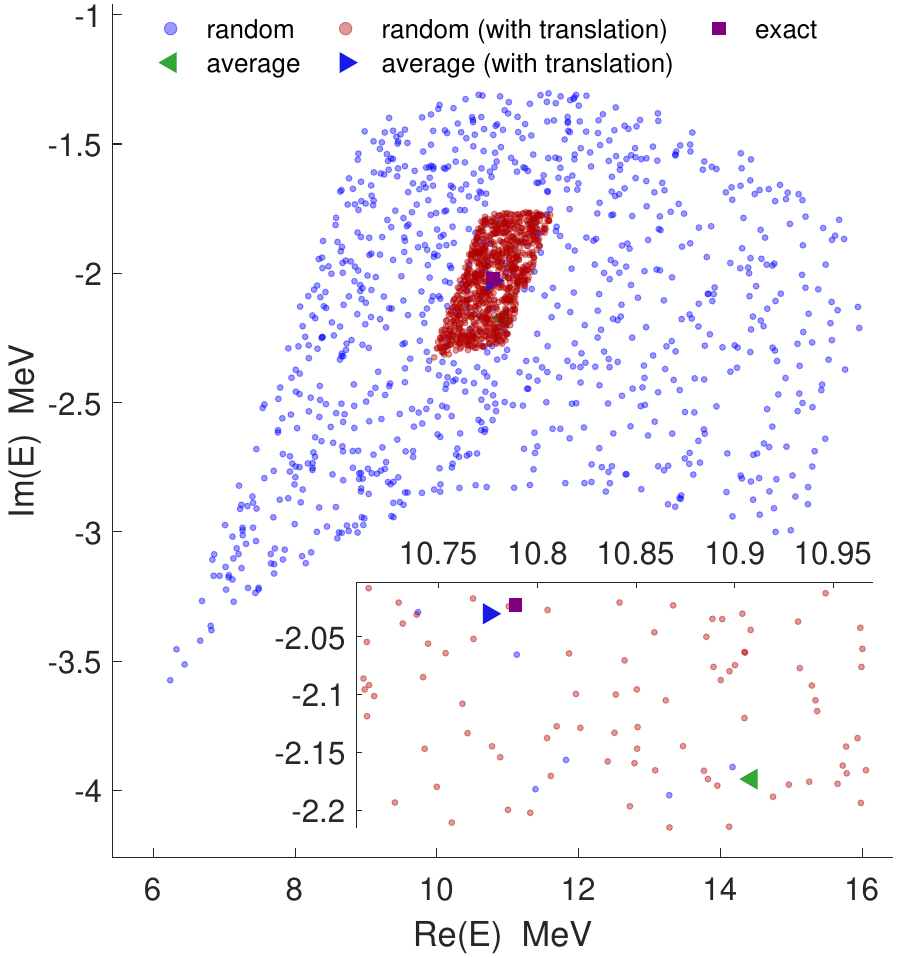} }
	\caption{
		Eigenvalues obtained after adding a random error ranging from -0.05 to 0.05 to the true values of possibilities  $p_1$ and  $p_2$.
		Here, the translation is taken as $t=\frac{Tr(C)}{4}$ and the global scaling is taken as $g=25-10i$. 	The blue points represent eigenvalues obtained with GS while the red points represent those with TGS. The purple squares represent the exact values, while the green left triangles and blue right triangles denote the average values of data points obtained using GS and TGS, respectively.}	\label{compare_translation} 
\end{figure}

  {\color{black} If we denote the eigenenergy $E$ as a function $E(p_1,p_2)$,  the total differential is $dE=\dfrac{\partial E}{\partial p_1}dp_1+\dfrac{\partial E}{\partial p_2}dp_2$, where $\dfrac{\partial E}{\partial p_i}$ are complex derivative values.  Thus, when $p_1$ and $ p_2$ take values within the small neighborhoods $[ p_1^0-dp_1, p_1^0 + dp_1]$ and $[ p_2^0-dp_2, p_2^0 + dp_2]$, the function $E(p_1, p_2)$ will fall within a parallelogram centered at $ E(p_1^0, p_2^0)$. If a sufficient number of random points are sampled for $p_1$ and $p_2$, then whether averaging eigenvalues directly or averaging possibilities first and then computing eigenvalues can both  represent the central value. If $dp_1$ and $dp_2$ are not small enough, the parallelogram may deform, but as long as the second and higher-order derivatives are relatively small, the average value remains meaningful and can provide a relatively reliable eigenvalue. Therefore, our goal can be summarized and described as making the first-order partial derivatives at the true values small (i.e., insensitivity to possibility and enhancing the reliability of single measurement) and avoiding the large deformation of the parallelogram (i.e., letting the higher-order derivatives be small and guaranteeing the effectiveness of averaging the eigenvalues).}  In this way, even when quantum computational errors are significant, the eigenenergy will still remain close to the true value without excessive dispersion. In practice, this allows us to obtain a more accurate and reliable eigenvalue by averaging the results obtained from quantum computing.

\begin{figure}[htbp] 
	\centering
	{\includegraphics[width=0.5\textwidth]{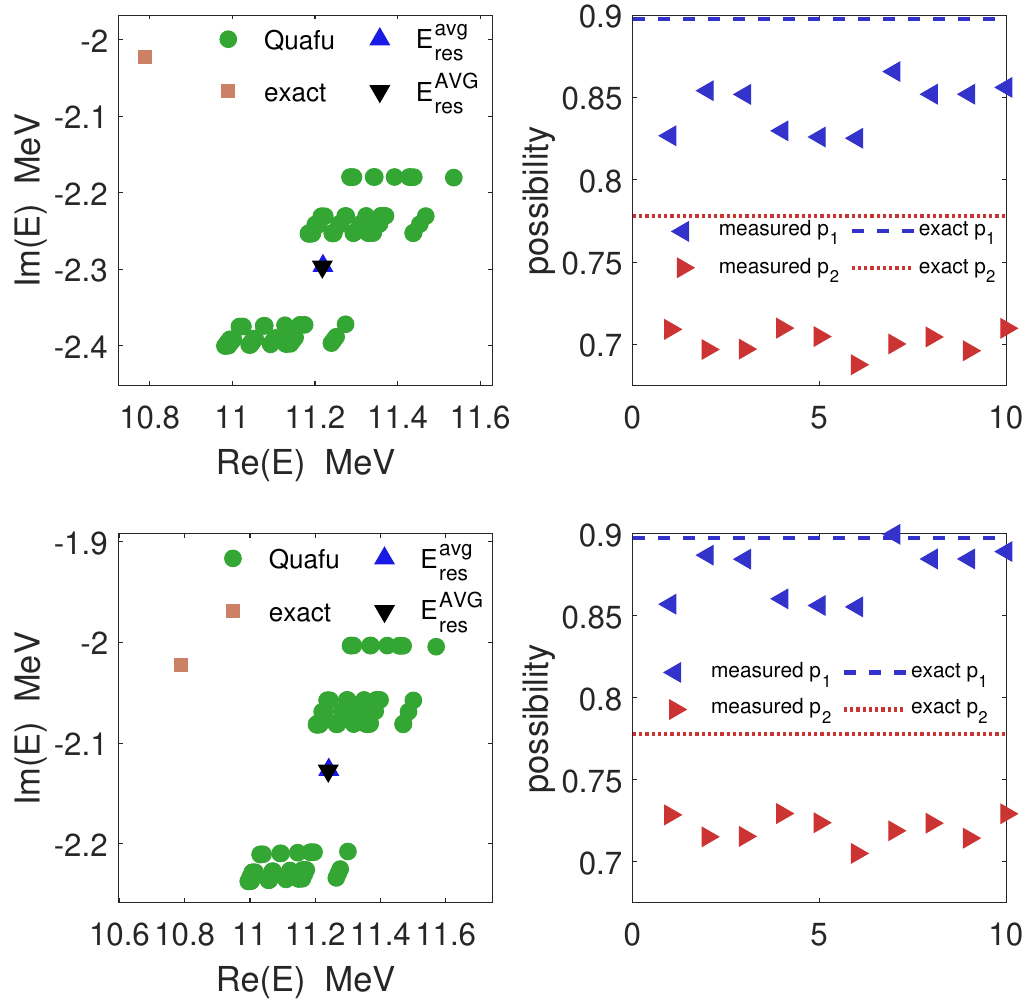} }
	\caption{The resonance state energies of two-body model obtained from quantum computer Quafu-Dongling with  $\theta=25.4343^{\circ}$, translation $t=\frac{Tr(C)}{4}$ and global scaling $g=25-10i$. The top two figures correspond to Dongling's original measurements, while the bottom two figures show the results after readout error mitigation. The green dots represent the  results obtained from Quafu-Dongling, the orange squares represent the exact values, and the blue upward triangles and black downward triangles correspond to the results from averaging possibilities and averaging all eigenvalues, respectively. The blue left triangles and red right triangles denote the measurement possibilities  $p_1$ and $p_2$, with the blue dashed and red dotted lines representing the corresponding exact values.}	
	\label{after_translation} 
\end{figure}

  However,  how can we achieve this? One feasible and straightforward approach is to apply a scaling to the original Hamiltonian $H\rightarrow g H$, where $g$ is a real or complex constant. In the following we call $g$ the global scaling of the original Hamiltonian. It is straightforward to prove that applying global scaling to the original Hamiltonian does not affect the value of possibility $p_1$, but only alters possibility $p_2$. Therefore, in Fig.\ref{FAS_oringinal_GS}, for ease of comparison, we use the same set of $p_1$ data points regardless of whether global scaling is applied.
 By choosing an appropriate global scaling, the eigenvalue's dependence on the measured possibilities can be reduced, thereby mitigating some of the quantum computation errors through averaging.  {\color{black}Explicitly speaking, in the problem we considered the sensitivity of the eigenvalues to possibility errors arises because the values of $\alpha_0$ and $\alpha_1$ calculated from the original Hamiltonian is relatively large. This means that even a small error in the possibility can significantly amplify the error in the final eigenvalue. To address this, we can consider using the global scaling approach to reduce the value of $\alpha$, namely choosing a global scaling greater than 1. Furthermore, we can introduce a complex global scaling, which might further reduce the dispersion of the results.
 	  As seen in the bottom left and right panels in Fig. \ref{FAS_oringinal_GS} when adopting the Hamiltonian after global scaling $g=25-10i$ the dispersion of the extracted eigenvalues become much smaller.  This allows us to achieve a more accurate and reliable eigenenergy.} 
\begin{figure}[htbp] 
	\centering
	{\includegraphics[width=0.5\textwidth]{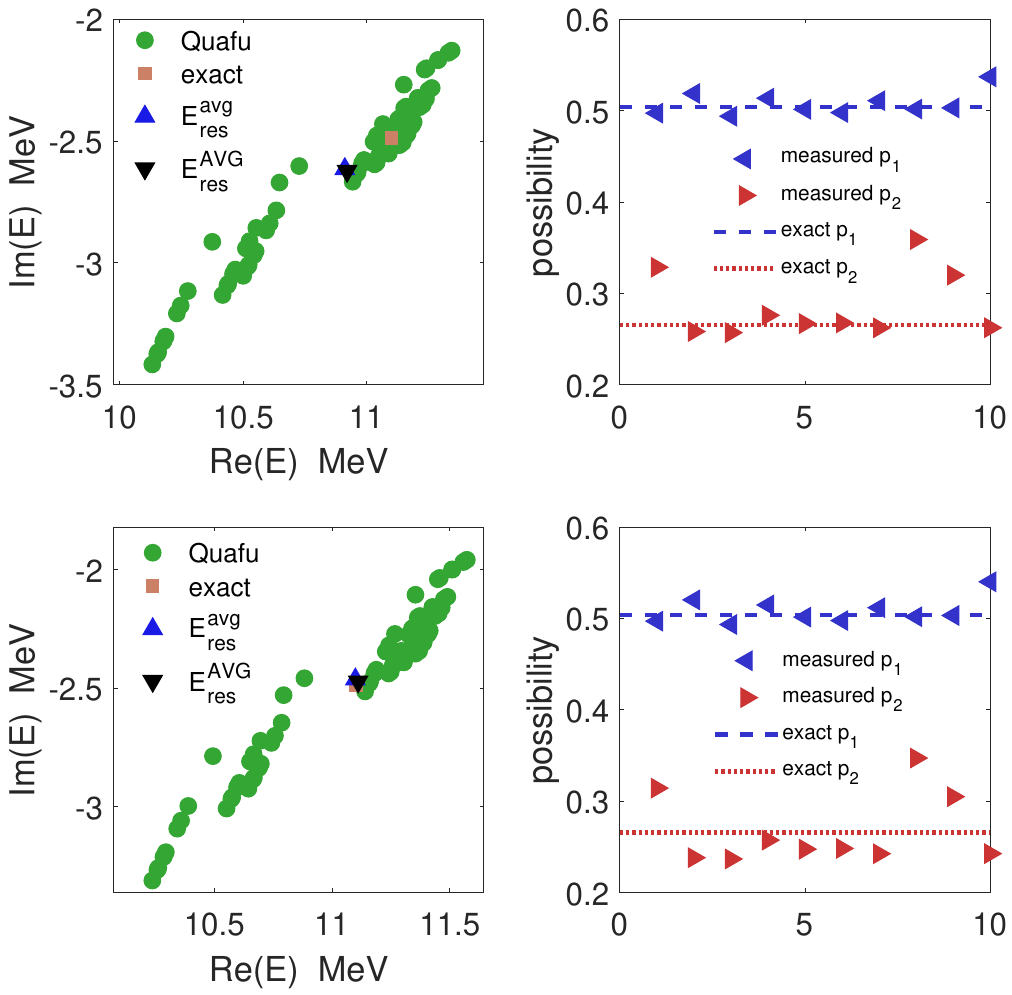} }
	\caption{The resonance state energies of THSR model obtained from quantum computer Quafu-Dongling with  $\theta=45.2356^{\circ}$, translation $t=\frac{Tr(C)}{4}$ and global scaling $g=10$. The top two figures correspond to Dongling's original measurements, while the bottom two figures show the results after readout error mitigation. The green dots represent the  results obtained from Quafu-Dongling, the orange squares represent the exact values, and the blue upward triangles and black downward triangles correspond to the results from averaging possibilities and averaging all eigenvalues, respectively. The blue left triangles and red right triangles denote the measurement possibilities  $p_1$ and $p_2$, with the blue dashed and red dotted lines representing the corresponding exact values.}	
	\label{THSR} 
\end{figure}

\begin{table*}
	\centering
	\begin{tabular}{lccccccc}
		\hline
		\multicolumn{8}{c}{$\alpha+\alpha$ with Gaussian interaction  (Model 1)  \ \ Direct diagonalization: 10.79-2.022i MeV}\\
		Approach&$p^{nr}_{1}$&$p_2^{nr}$&$E^{nr}_{res}$ (MeV) &$p^{avg}_{1}$&$p_2^{avg}$&$E^{avg}_{res}$ (MeV)&${E}_{res}^{AVG}$ (MeV)\\
		\midrule
		Full amplitude simulator (GS)&0.2986& 0.1117&$10.79-2.022i$&0.2974& 0.1111&$10.80-2.048i$&$10.80-2.051i$
		\\Quafu-Dongling (GS)&0.3001&0.1505&$ 8.378 - 2.241i$&0.3255&0.1682&$8.067 - 1.912i$&$8.074-1.953i$
		\\Quafu-Dongling (GS+REM)&0.2994&0.1106&$10.88 - 2.015i$&0.3036&0.1266&$9.899-2.003i$&$9.948-2.066i$\\
		Quafu-Dongling (TGS)&0.8658&0.7099&$ 11.28 - 2.179i$&0.8440&0.7016&$11.22 - 2.296i$&$11.22-2.297i$
		\\Quafu-Dongling (TGS+REM)&0.8998&0.7294&$11.31 - 2.003i$&0.8760&0.7204&$11.24-2.126i$&$11.24-2.127i$\\
		\hline
		\multicolumn{8}{c}{$\alpha+\alpha$ in microscopic THSR framework (Model 2)\ \ Direct diagonalization: 11.10-2.487i MeV}
		\\Quafu-Dongling (TGS)&0.5030&0.2670&$11.08 - 2.502i$&0.5075&0.2859&$10.91-2.614i$&$10.92-2.623i$
		\\Quafu-Dongling (TGS+REM)&0.5032&0.2575&$11.19 - 2.423i$&0.5082&0.2680&$11.10-2.463i$&$11.11-2.472i$\\
		\hline
	\end{tabular}
	\caption{
		The eigenvalues obtained through quantum simulation and quantum computation.  For $\alpha+\alpha$ with Gaussian interaction the exact result from direct diagonalization is $10.79-2.022i$ MeV and for THSR framework the  exact result from direct diagonalization is $11.10-2.487i$ MeV.  The second and third columns show the measured possibilities $p_1^{nr}$ and $p_2^{nr}$, which are nearest to the exact values $p_1^{exact}$ and $p_1^{exact}$, respectively. For Model 1,  $[p_1^{exact},p_2^{exact}]=[0.29905,0.11173]$ in the case of GS and $[0.89771,0.77818]$ in the case  of TGS, while for Model 2, $[p_1^{exact},p_2^{exact}]=[0.50352,0.26573]$ in the case of TGS. The fourth column shows the eigenvalue $E_{res}^{nr}$ obtained from $p_1^{nr}$ and $p_2^{nr}$. The fifth and sixth columns show the average values of $p^{avg}_1$ and $p^{avg}_2$, respectively. The seventh column lists the eigenvalue $E_{res}^{avg}$ obtained from the averaged $p^{avg}_1$ and $p^{avg}_2$. The final column presents the result ${E}_{res}^{AVG}$ obtained by averaging all the eigenvalues.}\label{tablesummary}
\end{table*}

 Based on these analyses and discussions, we proceed to compute the Hamiltonian after global scaling on the quantum computer Quafu-Dongling (8000 measurements). The corresponding results are displayed in Fig.\ref{quafuDongling}. It is shown that averaging yields a fairly accurate eigenvalue, which confirms the necessity of global scaling and validates the reliability of our approach. The top two figures in Fig.\ref{quafuDongling} show the original quantum computation results obtained by Dongling, while the bottom two figures display the results after readout error mitigation (REM). Here, the green dots represent the 10×10 sets of data obtained from quantum computation, the orange squares indicate the exact values, and the blue upward triangles and black downward triangles correspond to the results from averaging possibilities and averaging data points, respectively. The blue left triangles and red right triangles still denote the values of $p_1$ and $p_2$, with the blue dashed and red dotted lines representing the corresponding exact values. It is evident that even without REM, the original data can yield a reasonably reliable eigenvalue through averaging. After applying REM, the deviation of measurement possibilities from the true values is significantly reduced, and the error in the eigenvalue obtained by averaging is also notably decreased.

The role of global scaling can be summarized as scaling and rotating complex eigenvalues (the geometric meaning of complex multiplication). Naturally, one can think of another operation: translation, which means subtracting a constant multiple of the identity matrix from the original Hamiltonian matrix, namely $H \rightarrow H-t$. If we consider global scaling and translation as elements, and assume that complex multiplication and addition of eigenvalues do not change the quantum algorithm itself, then we obtain a group. It is easy to prove that the elements of this group can always be seen as applying translation first and then global scaling (or vice versa). However, it is important to note a special case: if this constant is half the trace of the matrix, both $p_1$ and $p_2$ will become exactly 1, making the quantum circuit completely trivial. As mentioned earlier, global scaling (GS) does not change the value of $p_1$, but if we combine the translation with global scaling (TGS), both possibilities can be controlled and potentially the dependence of eigenvalues on measurement possibilities can be further reduced. Then, based on the previously calculated example with only global scaling, we add a translation operation to optimize the robustness of the quantum computing.  Here, we first perform translation with $t=\dfrac{Tr(C)}{4}$ and then apply global scaling with $g=25-10i$. Using a stochastic simulation, we obtain the results as shown in Fig. \ref{compare_translation}. {\color{black} The blue points in Fig.\ref{compare_translation} represent results obtained using GS with random noise added in the range $[-0.05,0.05]$ to the exact possibility values. The red points show results obtained with the same random noise but using TGS. The purple squares represent the exact values, while the green left triangles and blue right triangles denote the average values of data points obtained using GS and TGS, respectively. It is evident that with TGS, the dispersion of the data is significantly reduced due to the additional degrees of freedom available to adjust the partial derivatives of the function $E(p_1,p_2)$. {\color{black} It is also noticeable that when using TGS for optimization, the consistency between averaging the eigenvalues and averaging the possibilities before computing the eigenvalue will be improved.} The corresponding results obtained from quantum computing are plotted in Fig. \ref{after_translation}. }

{\color{black}	
	As the final example, we consider the $\alpha+\alpha$ system within the microscopic THSR framework. The efficiency of the THSR wave function can significantly reduce the number of basis functions required for calculations, which is beneficial for quantum computing. Here, we use two THSR wave functions as basis vectors to calculate the resonance energy of the G-wave. The deformation parameters $[{\beta_z},\frac{\beta_{\bot}^2}{b^2}]$ used are $[10^{-6},0.3]$ and $[10^{-6},3]$ with $b$ being 1.36 fm. Through calculating the spectra the global optimal complex scaling angle is determined to be $45.2357^{\circ}$ and the eigenenergy is $11.100-2.487i$, which is pretty close to the accurate result $11.053-2.487i$ obtained with sufficient number of basis functions. Since the original  THSR wave functions are not orthonormal, we use Schmidt orthogonalization to construct new basis functions. In quantum computation,  the optimal complex scaling angle $45.2357^{\circ}$ is chose. As preliminary analysis shown, the robustness of the quantum computing can benefit from combining the global scaling with translation. In this example the parameters $t=\frac{Tr(C)}{4}$ and $g=10$ are chosen. As shown in Fig. \ref{THSR}, TGS yields a result with relatively small dispersion, and the average value is close to the exact eigenvalue.}

	Eventually,  we list the complex energies obtained from the various approaches including simulator and the quantum computer in Table \ref{tablesummary}. The second and third columns show the measured possibilities $p^{nr}_1$ and $p^{nr}_2$ closest to the exact values. The fourth column provides the eigenvalue $E_{res}^{nr}$ obtained from $p_1^{nr}$ and $p_2^{nr}$. The fifth and sixth columns contain the average values $p^{avg}_1$ and $p^{avg}_2$, respectively. The seventh column lists the eigenvalue $E_{res}^{avg}$ obtained from the averaged $p^{avg}_1$ and $p^{avg}_2$. The final column presents the result ${E}_{res}^{AVG}$ obtained by averaging all the eigenvalues.  As shown in Table \ref{tablesummary},  by introducing GS and TGS, the quantum computer can obtain reliable complex eigenvalues, and TGS effectively enhances the accuracy of the algorithm. In summary, considering the errors of quantum computers and the robustness of extracting complex resonant energy, the combination of global scaling and translation is useful to optimize the algorithm and obtain more reliable results. Moreover, TGS can significantly boost the confidence in a single quantum measurement thanks to smaller dispersion.

\section{conclusion}
\label{Conclusions}

We apply quantum computing to the complex scaled  Schrödinger equation and obtain the eigenenergies of the resonance states in  nuclear systems. For the case of two basis wave functions, we propose a new method for constructing quantum circuits. This method ensures that the measurement of ancilla qubit in the quantum circuit does not depend on the input eigenvector. Consequently, we can further simplify the quantum circuit, reduce the number of required quantum gates, and thereby lessen the impact of quantum computation errors, ultimately enhancing the stability of quantum computing. Moreover, this strategy for ensuring input eigenvector independence has the potential to be extended to cases with multiple basis vectors.

 In this work, we use the Quafu's quantum computer Dongling and the Origin quantum's full amplitude quantum simulator to achieve the computations. As an example, we calculate the G-wave resonant energy of $\alpha+\alpha$ system within both two-body toy model and microscopic nonlocalized cluster model. The results obtained from quantum computing are well consistent with those obtained from direct diagonalization.  Our proposed method can be applied not only to CSM for extracting resonance energies but also to the trap method for studying scattering properties under complex interactions \cite{Zhang:2024ykg}.  These are useful for using quantum computing to handle complex interaction systems in nuclear physics in the future.

\section*{Acknowledgements}
	This work is supported by the National Natural Science Foundation of China (Grants No.\ 12035011, No.\ 11905103, No.\ 11947211, No.\ 11761161001 , No.\ 11961141003, No.\ 12022517 and No.\ 12375122), by the National Key R\&D Program of China (Contracts No.\ 2023YFA1606503), by the Science and Technology Development Fund of Macau (Grants No.\ 0048/2020/A1 and No.\ 008/2017/AFJ), by the Fundamental Research Funds for the Central Universities (Grant No.\ 22120210138 and No.\ 22120200101).

\appendix

\section{}

If utilizing two basis functions (two system qubits) to obtain the Hamiltonian in second quantization form, one can find that it always can be expressed as,
\begin{equation}
	\begin{aligned}\label{twobasisH}
		&H_{\theta}=\beta_0 I\otimes I+\beta_1 I\otimes Z+\beta_2 Z\otimes I+\beta_3 (X\otimes X+Y\otimes Y)
	\end{aligned}
\end{equation}
where $\beta$ parameters are constants dependent on complex scale angle $\theta$,
\begin{equation}
	\begin{aligned}\label{}
		&\beta_0=\dfrac{C_{00}(\theta)+C_{11}(\theta)}{2},\ \beta_1=-\dfrac{C_{00}(\theta)}{2},\\&\beta_2=-\dfrac{C_{11}(\theta)}{2},\ \beta_3=\dfrac{C_{01}(\theta)+C_{10}(\theta)}{4}.
	\end{aligned}
\end{equation}

\begin{figure}[htbp] 
 	\centering
 	{\includegraphics[width=0.5\textwidth]{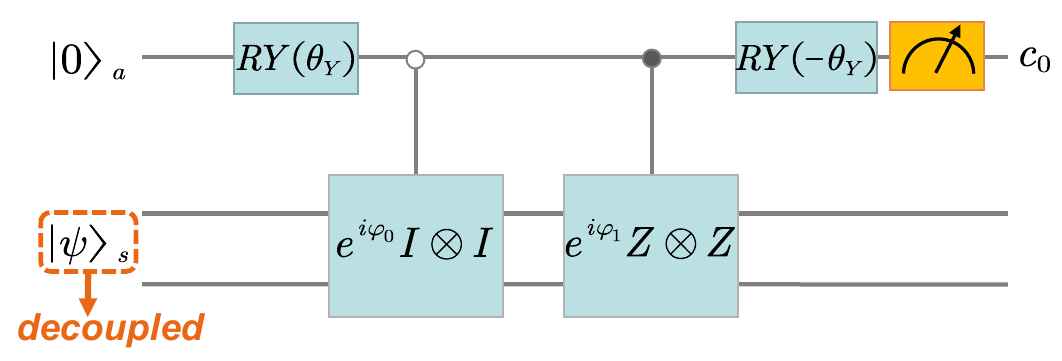} }
 	\caption{The quantum circuit for two basis wave functions and one ancilla qubit.  $A$ and $A^{\dagger}$ gates are realized through the rotation gate $RY(\theta_Y)$ and $RY^{\dagger}(\theta_Y)=RY(-\theta_Y)$.  The input eigenvector can be constructed through UCC, however, as we have mentioned, the measurement of the ancilla is not affected by input eigenvector $\ket{\psi}_s$ and therefore the input eigenvector can be eliminated.}	\label{specific_circuit} 
 \end{figure}

{\color{black}The figure  \ref{specific_circuit} shows a circuit consisting of two system qubits and one ancilla qubit. Notably, $A$ and $A^{\dagger}$can be implemented using either rotation gate $RY$ or $RX$.}

The matrix form of Eq.(\ref{twobasisH}) is written as,
\begin{equation}
	\begin{aligned}\label{}
		\begin{pmatrix}
			\beta_0+\beta_1+\beta_2& 0& 0& 0 \\  0& \beta_0-\beta_1+\beta_2& 2\beta_3& 0 \\  0 & 2\beta_3&\beta_0+\beta_1-\beta_2& 0 \\ 0 & 0 & 0& \beta_0-\beta_1-\beta_2
		\end{pmatrix}
	,
	\end{aligned}
\end{equation}
from which it can be conveniently proved that square of the new Hamiltonian $H_{\theta,new}^{}=H_{\theta}-\beta_0I\otimes I$ has the form of,
\begin{equation}
	\begin{aligned}\label{}
		&H_{\theta,new}^{2}=\gamma_0I\otimes I+\gamma_1Z\otimes Z
	\end{aligned}
\end{equation}
where coefficients $\gamma_0$ and $\gamma_1$ are given by,
\begin{equation}
	\begin{aligned}\label{}
		&\gamma_0=\dfrac{C_{00}(\theta)^2+C_{11}(\theta)^2+2C_{01}(\theta)C_{10}(\theta)}{4}\\&\gamma_1=\dfrac{C_{00}(\theta)C_{11}(\theta)-C_{01}(\theta)C_{10}(\theta)}{2}
	\end{aligned}
\end{equation}

Additionally, one can also prove that any non-negative even power of $H_{\theta,new}^{}$ can be written as a linear combination of operators $I\otimes I$ and $Z\otimes Z$, in other words, the even powers of $H_{\theta,new}$ form a group. Therefore, in order to determine the magnitude and argument of the complex eigenvalue we can choose operators $H_{\theta,new}^{2}$ and $H_{\theta,new}^{2}+H_{\theta,new}^{4}$ and make measurement on ancilla qubit. Such choice is exactly what we adopt in this work.  Finally, the complex eigenvalues are written as,
\begin{equation}
	\begin{aligned}\label{pdfdetermine}
		& E=\beta_0 \pm p^{1/4}A^{1/2}\\&\exp(\dfrac{\rm{sign}(\mathcal{I}(\gamma_1-\gamma_0))}{2i}\cos^{-1}(\dfrac{p'^2A'^2}{2p^{3/2}A^3}-\dfrac{1}{2p^{1/2}A}-\dfrac{p^{1/2}A}{2}))
	\end{aligned}
\end{equation}
where  $\mathcal{I}$ denotes taking  the imaginary part. Certainly, we can also use other methods to obtain the eigenvalues, such as forming other linear combinations of $H_{\theta,new}^{2}$ and $H_{\theta,new}^{4}$, or using different operators like $H_{\theta,new}^{6}$, $H_{\theta,new}^{8}$, etc. This will yield expressions that are different in form but equivalent to Eq.(\ref{pdfdetermine}).

Generally, the input of the system qubit needs to be an specific eigenvector. However, through derivation using the above framework, we can find that the eigenvector and the measurement of ancilla qubit is decoupled. This means the measurement possibility of the ancilla does not depend on the input eigenvector, which provides a universal method to solve the eigenvalues of a second-order complex symmetric matrix. Consequently, we can simplify the original three-qubit circuit to a two-qubit circuit with only five quantum gates.  This can effectively improve the accuracy of quantum circuit operations and reduce the impact of errors and noise.

If considering the original Hamiltonian matrix,
\begin{equation}
	\begin{aligned}\label{ordertwo}
		&C\psi=\begin{pmatrix}
			C_{00}\  C_{01}\\C_{10}\  C_{11}
		\end{pmatrix}\psi
		=E\psi
	\end{aligned}
\end{equation}
we can easily obtain the eigenvalues $E$ as,
\begin{equation}
	\begin{aligned}\label{}
		&E=\dfrac{C_{00}+C_{11}\pm\sqrt{(C_{00}-C_{11})^2+4C_{01}C_{10}}}{2}
	\end{aligned}
\end{equation}

By comparing  Eq.(\ref{pdfdetermine}) and Eq.(\ref{ordertwo}), it is clear that we have obtained a form for quantum computing that is equivalent to diagonalize the Hamiltonian matrix.

\bibliographystyle{elsarticle-num-names} 
\nocite{*}
\bibliography{example}

\end{document}